\documentclass[fleqn,usenatbib]{mnras}

\usepackage{newtxtext,newtxmath}

\usepackage[T1]{fontenc}

\DeclareRobustCommand{\VAN}[3]{#2}
\let\VANthebibliography\thebibliography
\def\thebibliography{\DeclareRobustCommand{\VAN}[3]{##3}\VANthebibliography}

\usepackage{graphicx}	
\usepackage{amsmath}	
\usepackage{orcidlink}

\title[Young pulsars from white dwarf mergers]{\vspace{-0.75cm} Connecting the Young Pulsars in Milky Way Globular Clusters with White Dwarf Mergers and the M81 Fast Radio Burst}

\author[Kremer et al.]{
\parbox[t]{\textwidth}{\vspace{-0.75cm}
Kyle Kremer \orcidlink{0000-0002-4086-3180},$^{1}$\thanks{NASA Einstein Fellow}
Jim Fuller \orcidlink{0000-0002-4544-0750},$^1$
Anthony L. Piro \orcidlink{0000-0001-6806-0673},$^{2}$
and Scott M. Ransom \orcidlink{0000-0001-5799-9714}$^3$} \vspace{-0.25cm} \\
$^{1}$ TAPIR, California Institute of Technology, Pasadena, CA 91125, USA. E-mail: kkremer@caltech.edu\\
$^{2}$The Observatories of the Carnegie Institution for Science, Pasadena, CA 91101, USA\\
$^3$NRAO, 520 Edgemont Road, Charlottesville, VA 22903, USA
\vspace{-0.5cm}
}

\date{\vspace{-0.5cm}\today}
\pubyear{2023}
\begin{document}
\maketitle

\begin{abstract}
The detections of four apparently young radio pulsars in the Milky Way globular clusters are difficult to reconcile with standard neutron star formation scenarios associated with massive star evolution. Here we discuss formation of these young pulsars through white dwarf mergers in dynamically-old clusters that have undergone core collapse. Based on observed properties of magnetic white dwarfs, we argue neutron stars formed via white dwarf merger are born with spin periods of roughly $10-100\,$ms and magnetic fields of roughly $10^{11}-10^{13}\,$G. As these neutron stars spin down via magnetic dipole radiation, they naturally reproduce the four observed young pulsars in the Milky Way clusters. Rates inferred from $N$-body cluster simulations as well as the binarity, host cluster properties, and cluster offsets observed for these young pulsars hint further at a white dwarf merger origin. These young pulsars may be descendants of neutron stars capable of powering fast radio bursts analogous to the bursts observed recently in a globular cluster in M81.
\end{abstract}

\begin{keywords} globular clusters: general -- pulsars:general -- stars:white dwarf -- transients: fast radio bursts \vspace{-0.75cm}
\end{keywords}



\vspace{-1cm}
\section{Introduction}

The globular clusters in the Milky Way are known to host robust populations of radio pulsars \citep[e.g.,][]{Lyne1987, psr_catalog}. 
For any typical initial mass function \citep[e.g.,][]{Kroupa2001}, it is well-understood that large numbers of neutron stars form at early times ($t\lesssim50\,$Myr) in a typical cluster through standard iron core-collapse and/or electron-capture supernovae \citep[e.g.,][]{Podsiadlowski2004}, the latter of which may be necessary to produce sufficiently small natal kicks to enable retention in the relatively shallow potential wells of typical globular clusters \citep[e.g.,][]{Pfahl2002}. However, in present-day globular clusters more than $10\,$Gyr old, pulsars formed through massive stellar evolution will have long ago spun down through magnetic dipole radiation rendering them undetectable as radio sources \citep[e.g.,][]{RudermanSutherland1975}. Thus additional processes are necessary to explain the plethora of radio pulsars seen in old globular clusters.

One method is the classic low-mass X-ray binary scenario where the neutron star is ``recycled'' and spun up to millisecond spin periods through accretion from a binary companion \citep[e.g.,][]{Alpar1982}. Indeed, the majority of observed globular cluster pulsars are so-called millisecond pulsars and the processes through which such systems may form dynamically in a dense stellar environment have been well-studied \citep[e.g.,][]{SigurdssonPhinney1995,Ivanova2008,Ye2019}. Furthermore, this scenario connects naturally with the well-known overabundance of X-ray sources in globular clusters relative to the Galactic field \citep[e.g.,][]{Clark1975}. 

However, alongside the broader class of presumably recycled radio pulsars in globular clusters, there is an additional class of four apparently young pulsars \citep{Lyne1996,Boyles2011} with relatively long spin periods of $0.1-1\,$s and inferred magnetic fields of roughly $10^{11}-10^{12}\,$G, larger than those expected in a recycling scenario where accretion is likely to ``bury'' any residual neutron star magnetic field \citep[e.g.,][]{BhattacharyaVanDenHeuvel1991}. These four pulsars, which have characteristic ages of roughly $10^7-10^8\,$yr (see Table~\ref{table:pulsars} for a summary of observed properties), have been touted as evidence for alternative formation scenarios, in particular involving collapse of massive white dwarfs \citep[e.g.,][]{Tauris2013}. 
Such white dwarf dynamics are motivated by observed populations of cataclysmic variables in globular clusters \citep[e.g.,][]{Grindlay1995} and also by $N$-body simulations of white dwarfs in clusters that naturally lead to both accretion-induced collapse in binaries and massive white dwarf mergers \citep[e.g.,][]{Kremer2021_wd}.

\begin{table*}
	\centering
    \renewcommand{\arraystretch}{1}
    \tabcolsep=10pt
	\caption{Properties of the four young pulsars observed in Milky Way globular clusters adapted from \citet{Boyles2011}.}
	\label{table:pulsars}
	\begin{tabular}{l|cccccccc} 
\hline
\hline
\multicolumn{1}{c}{Pulsar} &
\multicolumn{1}{c}{Host Cluster} &
\multicolumn{1}{c}{$P$} &
\multicolumn{1}{c}{$\dot{P}$} &
\multicolumn{1}{c}{$B$} &
\multicolumn{1}{c}{Age} &
\multicolumn{1}{c}{Cluster Offset} &
\multicolumn{1}{c}{Binary?} &
\multicolumn{1}{c}{Reference} \\
\multicolumn{1}{c}{} &
\multicolumn{1}{c}{} &
\multicolumn{1}{c}{(ms)} &
\multicolumn{1}{c}{($\rm{s\,s}^{-1}$)} &
\multicolumn{1}{c}{(G)} &
\multicolumn{1}{c}{(yr)} &
\multicolumn{1}{c}{(pc)} &
\multicolumn{1}{c}{} &
\multicolumn{1}{c}{} \\
\hline
PSR B1745-20 & NGC 6440 & 288 & $4.0\times10^{-16}$ & $3.4\times10^{11}$ & $1.1\times10^7$ & 0.12 & No & (a) \\
PSR J1823-3031B & NGC 6624 & 379 & $3.0\times10^{-17}$ & $1.1\times10^{11}$ & $2.0\times10^8$ & 0.53 & No & (b) \\
PSR J1823-3021C & NGC 6624 & 406 & $2.2\times10^{-16}$ & $3.0\times10^{11}$ & $2.9\times10^7$ & 0.31 & No & (b) \\
PSR B1718-19 & NGC 6342 & 1004 & $1.6\times10^{-15}$ & $1.3\times10^{12}$ & $9.8\times10^6$ & 5.7 & Yes & (c) \\
	\hline
    \hline
    \multicolumn{9}{|p{\linewidth}|}{\textit{References:} (a) \citet{Freire2008}; (b) \citet{Lynch2012}; (c) \citet{Lyne1996}  }
	\end{tabular}
\end{table*}

A recently-observed repeating fast radio burst (FRB) localized to an old globular cluster in M81 \citep{Bhardwaj2021,Kirsten2022} added a new piece to this puzzle. The popular core-collapse-supernova magnetar mechanism for FRB sources \citep[e.g.,][]{PopovPostnov2013,Bochenek2020} is clearly inconsistent with this cluster source; any magnetars formed through standard massive stellar evolution in the cluster will have been inactive for billions of years by the present day. Alternatively, recent studies \citep{Kremer2021_FRB,Lu2022} have argued this source may instead be powered by a neutron star born recently through collapse of a massive white dwarf, similar to the channel through which the aforementioned four young pulsars may have formed.

In this Letter, we discuss the formation of the four young radio pulsars in the Milky Way globular clusters via collapse of white dwarf merger remnants and connect these objects with the M81 FRB source. In Section~\ref{sec:NSs}, we describe the characteristic birth properties of neutron stars formed through this scenario, motivated by properties of isolated magnetic white dwarfs. 
In Section~\ref{sec:further} we describe additional constraints that point toward a white dwarf merger origin, namely host cluster properties, the binarity (or lack thereof) of the four young pulsars, and the host cluster offsets. In Section~\ref{sec:FRB} we connect these objects with the M81 FRB repeater and argue the observed young pulsars may be descendants of FRB sources. We summarize and conclude in Section~\ref{sec:conclusion}.

\vspace{-0.7cm}
\section{Young neutron stars from white dwarf mergers in globular clusters}
\label{sec:NSs}

In globular clusters, compact objects play an essential role in the evolution of their host environment \citep[e.g.,][]{Mackey2007,BreenHeggie2013}. At early times ($t\lesssim10\,$Gyr), stellar black holes dynamically ``heat'' their host cluster through encounters with one another and with cluster stars \citep[e.g.,][]{kremer2020modeling}. Through these encounters, black holes are ``kicked'' through gravitational recoils and are ultimately ejected from their host; thus the number of black holes decreases as the cluster evolves \citep[e.g.,][]{Kulkarni1993}. At late times ($t\gtrsim10\,$Gyr), once nearly all black holes have been ejected, some clusters can undergo core collapse \citep[e.g.,][]{kremer2019initial}. At this point, the next most massive cluster objects -- massive white dwarfs and neutron stars -- form their own dense central subsystem \citep[e.g.,][]{Vitral2022}. Within these ultra-dense ($n\gtrsim10^6\,\rm{pc}^{-3}$) centres of core-collapsed clusters, white dwarf binaries form dynamically leading to high rates of white dwarf binary mergers. Using $N$-body cluster simulations incorporating these various processes relevant to compact object dynamics, \citet{Kremer2023_FRB} estimated a white dwarf merger rate of roughly $10^{-9}\,\rm{yr}^{-1}$ per typical cluster. As a consequence of mass segregation, the vast majority \citep[$\gtrsim90\%$;][]{Kremer2021_wd} of these white dwarf merger pairs have a total mass in excess of the Chandrasekhar limit suggesting most lead to collapse and formation of young neutron stars \citep[e.g.,][]{NomotoIben1985,Schwab2021}. These relatively high masses may yield an overabundance of neutron stars formed via this channel in clusters relative to mergers in the galactic field, which are preferentially lower mass albeit at a higher overall rate \citep[e.g.,][]{YungelsonLivio1998}.

The properties of neutron stars born through collapse of white dwarf merger remnants are uncertain and depend on complex evolution of angular momentum and magnetic field during the merger \citep[e.g.,][]{Dan2014}, subsequent stellar evolution \citep[e.g.,][]{Schwab2021}, and ultimately, the collapse itself \citep[e.g.,][]{King2001,Dessart2006}. However, some basic insight can be obtained from the observed properties of isolated highly-magnetic white dwarfs \citep[e.g.,][]{Ferrario2015}, some of which may have formed through mergers of lower-mass white dwarfs with total mass below the Chandrasekhar limit \citep[e.g.,][]{Caiazzo2021}.

Consider a white dwarf merger remnant with properties comparable to the isolated magnetic white dwarfs of \citet{Ferrario2015} but with total mass over the Chandrasehkar limit so that the merger results in collapse.\footnote{The merger candidates in \citet{Ferrario2015} are all far from the Chandrasehkar limit, however more massive merger remnants exhibit similar spin periods and field strengths (Ilaria Caiazzo, private communication).} Assuming conservation of angular momentum and magnetic flux of the observed white dwarf sample during such collapse, we infer characteristic values for the magnetic field strength and spin period of these hypothetical neutron stars at birth:

\begin{multline}
    B_{\rm{ns}}=B_{\rm{wd}} (R_{\rm{wd}}/R_{\rm{ns}})^2 \\\shoveright{P_{\rm{ns}}=P_{\rm{wd}} (R_{\rm{wd}}/R_{\rm{ns}})^{-2}.}
\end{multline}

White dwarf magnetic fields of roughly $10^6-10^8\,$G and spin periods of roughly $1-10\,$hr \citep[e.g.,][]{Ferrario2015} imply neutron star field strengths of roughly $10^{11}-10^{13}\,$G
and spin periods of $10-100\,$ms \cite[assuming $R_{\rm ns}=10^6\,$cm and $R_{\rm wd}\approx3\times10^8\,$cm;][]{Caiazzo2021}. 
Here we have glossed over the post-merger details which include phases of viscous and thermal evolution (potentially including dusty mass loss that in turn affects angular momentum evolution) as long as $\sim10^4\,$yr before the ultimate collapse \citep[e.g.,][]{Shen2012,Schwab2016,Schwab2021}. However, these more detailed studies generally predict neutron star properties consistent with our order-of-magnitude estimates, as long as angular momentum losses due to winds are not too high.

\begin{figure}
    \includegraphics[width=0.95\columnwidth]{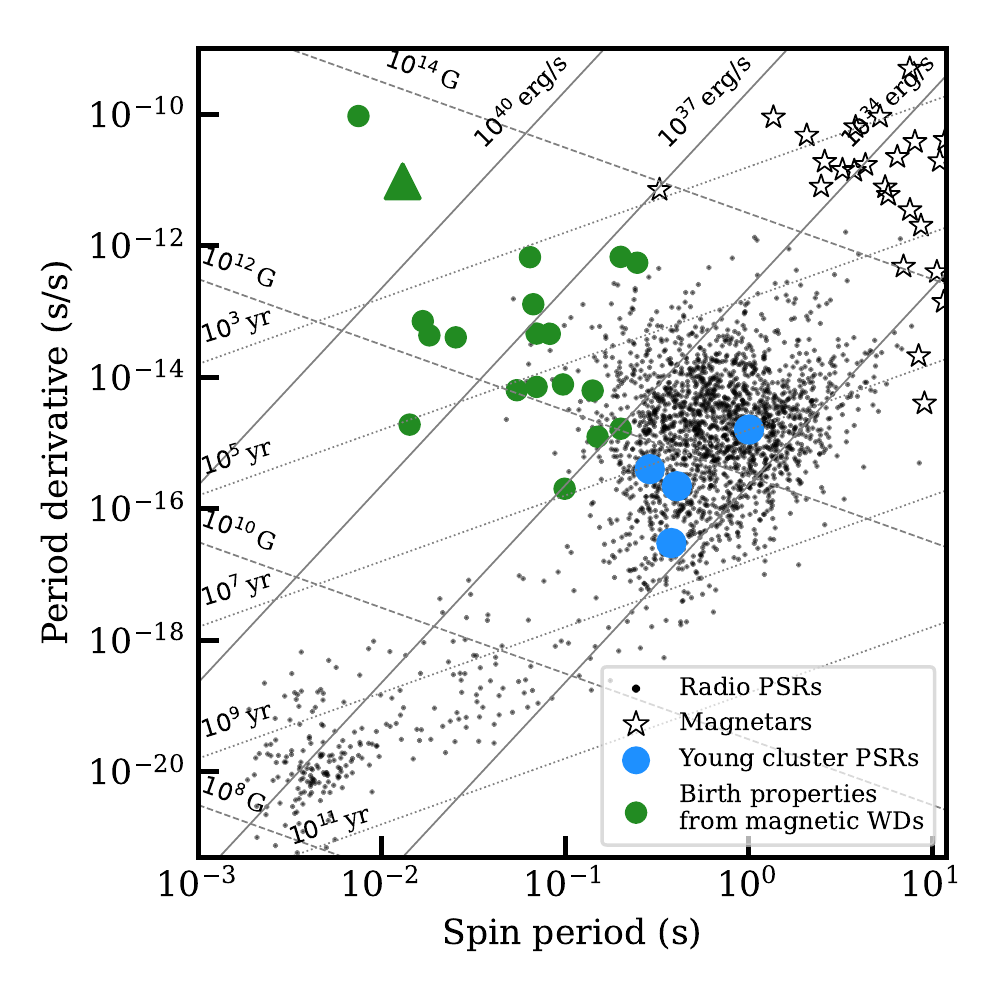} \vspace{-0.2cm}
    \caption{ $P\dot{P}$ diagram. In black we show all observed Galactic radio pulsars \citep[circles;][]{Manchester2005} and Galactic magnetars \citep[stars;][]{OlausenKaspi2014}. In blue we show the four apparently young pulsars in the Milky Way globular clusters \citep{Boyles2011}. In green, we show estimates for birth properties of neutron stars formed via white dwarf mergers computed from the observed features of isolated white dwarfs in \citet{Ferrario2015} and (the green triangle) \citet{Caiazzo2021}. As the neutron stars represented as green points evolve along tracks of constant magnetic field (dashed contour lines) as they spin down via magnetic dipole radiation, they pass through the $P\dot{P}$ region occupied by the four young pulsars, suggesting a connection between these two populations. \vspace{-0.5cm}}
    \label{fig:PPdot}
\end{figure}

In Figure~\ref{fig:PPdot} we show the locations of these objects in a $P\dot{P}$ diagram. Green symbols show locations of neutron stars formed through massive white dwarf mergers, with properties inferred from conservation of angular momentum and magnetic flux of the highly-spinning ($P < 10\,$hr) and highly-magnetic ($B>10^6\,$G) single white dwarfs in the \citet{Ferrario2015} sample (green circles) and ZTF J190132.9+145808.7 (a likely white merger product) from \citet{Caiazzo2021} (green triangle).\footnote{These spin periods may be viewed as upper limits, since in reality these observed white dwarfs may have spun down from their initial values due to magnetic winds \citep[e.g.,][]{Gvaramadze2019}
.} Blue symbols show the four young pulsars in the Milky Way globular clusters of Table~\ref{table:pulsars}.

Once formed, a neutron star evolving via magnetic dipole radiation will spin down over time, following tracks of roughly constant characteristic field strength (dotted lines in Figure~\ref{fig:PPdot})

\begin{equation}
    \label{eq:B}
    B \approx \Bigg( \frac{3c^3I}{8\pi^2R_{\rm ns}^6} \Bigg)^{1/2} (P \dot{P} )^{1/2} \approx 10^{12} \Bigg( \frac{P}{100\,\rm{ms}}\Bigg)^{1/2} \Bigg( \frac{\dot{P}}{10^{-14}\rm{s/s}}\Bigg)^{1/2}\,\rm{G}
\end{equation}
(here $I\approx 0.4 M R_{\rm ns}^2$ is the neutron star moment of inertia) and crossing lines of constant characteristic age (dashed lines in Figure~\ref{fig:PPdot})

\begin{equation}
    \label{eq:tau}
    \tau \approx \frac{P}{2\dot{P}} \approx 10^5\, \Bigg( \frac{\it{P}}{50\,\rm{ms}}\Bigg)^2 \Bigg( \frac{\it{B}}{10^{12}\,\rm{G}}\Bigg)^{-2}\,\rm{yr}
\end{equation}
\citep[e.g.,][]{ShapiroTeukolsky1983}. As shown in Figure~\ref{fig:PPdot}, as young neutron stars formed through white mergers spin down, they pass through the region occupied by the four young cluster pulsars, hinting at an evolutionary connection between these populations.

Eventually, pulsars spin down sufficiently to fall below the so-called ``death line'', an empirical boundary near $B/P^2 = 1.7\times10^{11}\,\rm{G\,s}^{-2}$ \citep[e.g.,][]{RudermanSutherland1975} below which they no longer emit observable radiation. For a given initial magnetic field, the characteristic spin-down time to reach this death line is roughly $\tau_{\rm sd} \approx 10^8 (B/10^{12}\,\rm{G})^{-1}\,\rm{yr}$. Consider the Milky Way which contains roughly 100 globular clusters \citep{Harris1996}. If neutron stars are formed via white dwarf mergers at a rate  $\mathcal{R}\approx 10^{-9}\,\rm{yr}^{-1}$ per cluster \citep{Kremer2023_FRB}, this implies a neutron star formation rate of roughly $10^{-7}\,\rm{yr}^{-1}$ in the full Milky Way cluster population. For a given magnetic field strength at formation (presumably roughly in the range $10^{11}-10^{13}\,$G motivated by the green points in Figure~\ref{fig:PPdot}) corresponding to a spin-down time $\tau_{\rm sd}$, we estimate

\begin{equation}
    N_{\rm obs} \approx 10\,\Bigg( \frac{B}{10^{12}\,\rm{G}} \Bigg)^{-1} \Bigg( \frac{N_{\rm cl}}{100} \Bigg) \Bigg(\frac{\mathcal{R}}{10^{-9}\,\rm{yr}^{-1}\rm{cl}^{-1}}\Bigg)
\end{equation}
apparently young pulsars in the Milky Way globular clusters today, comparable to the population of four such objects observed.

\vspace{-0.5cm}
\section{Further considerations}
\label{sec:further}

\subsection{Host cluster properties}

A key prediction from $N$-body simulations is that white dwarf mergers occur by far most frequently in dynamically-evolved clusters that have ejected nearly all of their stellar-mass black holes \citep[e.g.,][]{kremer2020modeling}. In the absence of dynamical energy generated through black hole binary dynamics, the core of the cluster begins to collapse \citep[for this reason, core radius is a reasonable proxy for dynamical age;][]{Spitzer1969}. As a consequence, white dwarf mergers occur most often (by factors $\gtrsim100$) in clusters that have reached, or have nearly reached, core collapse. If indeed our four observed young pulsars are formed via white dwarf mergers, we naturally expect the pulsars to live in host clusters of this type.

NGC~6624 (which contains \textit{two} young pulsars) and NGC~6342 are both core-collapsed \citep{Harris1996} and, although not traditionally categorized as core-collapsed, NGC~6440 is indeed a very dynamically-evolved cluster with a small core \citep[e.g.,][]{Pallanca2021}. In Figure~\ref{fig:rcrh} we show observed values for core radius versus half-light radius for all globular clusters in the Milky Way \citep[data from][]{Harris1996}. Filled and open circles show clusters that have and have not undergone cluster core collapse. Blue points denote the cluster hosts of the four young pulsars, all of which populate the region of parameter space occupied by the most dynamically-evolved clusters. Thus our expectation for finding young pulsars in the densest and dynamically oldest clusters is met.

\vspace{-0.4cm}
\subsection{Singles versus binaries}

Since the white dwarf companion is disrupted entirely during a white dwarf merger \citep[e.g.,][]{Dan2014}, neutron stars formed through collapse of white dwarf merger remnants are naturally expected to be single objects. Notably, three of the four young globular cluster pulsars are singles, consistent with this expectation. The exception is PSR BJ1718-19, an eclipsing binary with a roughly $6\,$hr orbit and companion mass $\geq0.12\,M_{\odot}$ \citep{Lyne1993}. Previous analyses \citep[e.g.,][]{SigurdssonPhinney1995} have shown such systems may form through binary exchange interactions after neutron star formation. Furthermore, $N$-body simulations of core-collapsed clusters generally predict binary fractions of roughly $5-10\%$ for objects of this type \citep[e.g.,][]{Ye2019,kremer2020modeling}, consistent with a binary fraction of $1/4$. Additionally, we note the caveat that the association of PSR BJ1718-19 with its presumed host cluster NGC~6342 is the subject of some doubt \citep[see discussion in][]{Boyles2011}. If in fact this object is not truly a member of its host cluster, then all of the young pulsars in globular clusters are single objects.

\vspace{-0.4cm}
\subsection{Cluster offsets}
\label{sec:offsets}

\begin{figure}
    \includegraphics[width=0.95\columnwidth]{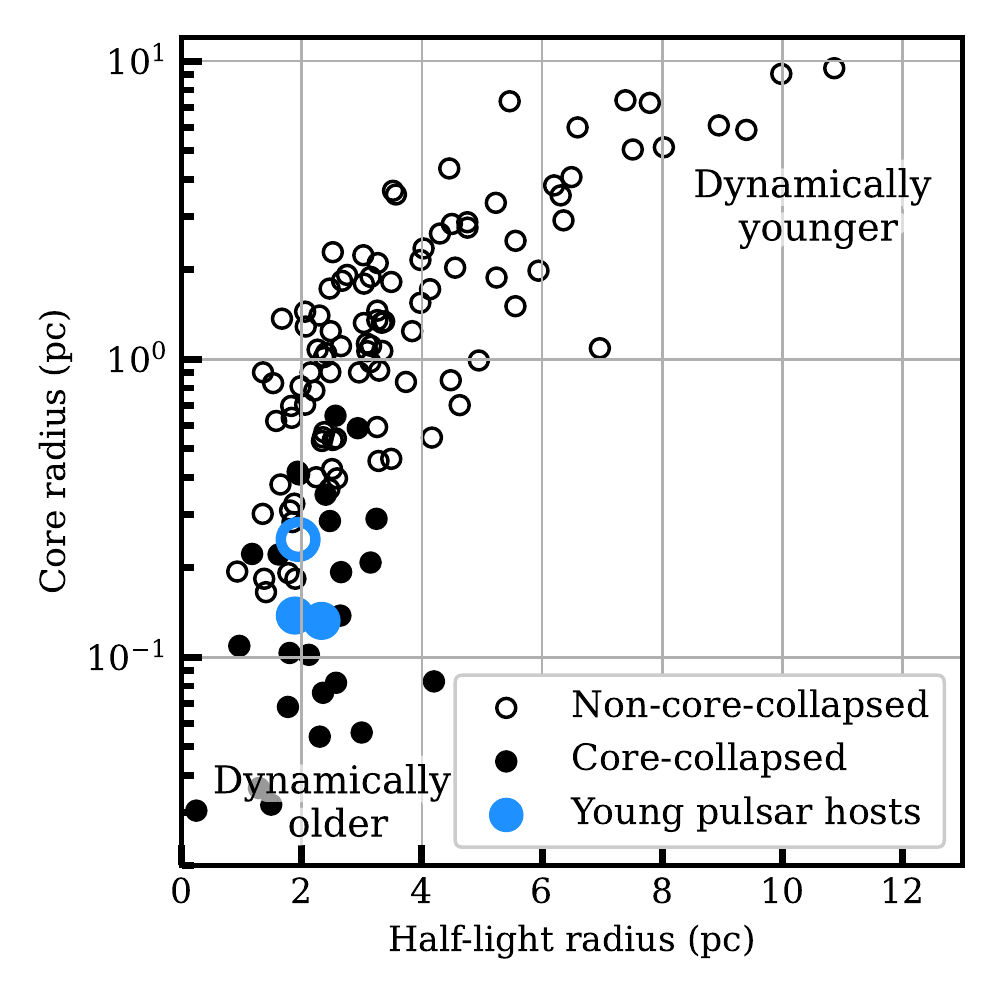} \vspace{-0.1cm}
    \caption{Core versus half-light radius for all Milky Way globular clusters. Data taken from \citet{Harris1996} and \citet{Pallanca2021} for NGC~6440. Filled (open) circles denote clusters that have (have not) reached core collapse. Blue symbols denote the three cluster hosts of the four young radio pulsars. As shown, the hosts for all four of these pulsars are among the densest and most dynamically-evolved globular clusters in the Milky Way, consistent with expectations for a white dwarf merger origin \citep{Kremer2021_wd}. \vspace{-0.3cm}}
    \label{fig:rcrh}
\end{figure}

The projected offsets of the young pulsars also yield insight into their origin. Three of the four objects are found well within the half-light radii (roughly $2\,$pc; see Figure~\ref{fig:rcrh}) of their host clusters. This is consistent with expectations from mass segregation, even accounting for natal kicks of order $10\,$km/s that may initially place these neutron stars on wide cluster orbits \citep[e.g.,][]{Kremer2023_FRB}.

The mass segregation timescale for neutron stars of mass $m\approx 1.5\,M_{\odot}$ is roughly $t_{\rm ms} \sim (m_{\rm avg}/m)t_{\rm rh}$ where $m_{\rm avg} \approx 0.5\,M_{\odot}$ is the average mass of all cluster objects and $t_{\rm rh}$ is the half-mass relaxation time \citep[e.g.,][]{Spitzer1969}

\begin{multline}
    t_{\rm rh} = \frac{0.14 M_{\rm cl}^{1/2} r_h^{3/2}}{G^{1/2} m_{\rm avg} \ln \Lambda} \\ \approx 10^8 \Bigg( \frac{M_{\rm cl}}{2\times10^5\,M_{\odot}} \Bigg)^{1/2} \Bigg( \frac{r_h}{2\,\rm{pc}} \Bigg)^{3/2} \Bigg( \frac{m_{\rm avg}}{0.5\,M_{\odot}} \Bigg)^{-1} \Bigg( \frac{\ln \Lambda}{10} \Bigg)^{-1}\,\rm{yr},
\end{multline}
where we have adopted Coulomb logarithm $\ln \Lambda =10$, appropriate for the clusters of interest here. Thus the young pulsars with characteristic ages of roughly $10^7-10^8\,$yr will quite reasonably have had sufficient time to mass segregate to their hosts' centres.

The exception again is the binary pulsar BJ1718-19 which is offset by nearly $6\,$pc, well outside its host's half-mass radius. Keeping in mind the previous caveat about cluster membership, it is possible that such an offset may be explained through gravitational recoil associated with a binary exchange encounter that may have formed this system \citep[e.g.,][]{Heggie1975,SigurdssonPhinney1995}.

\section{Connecting to the M81 Fast Radio Burst}
\label{sec:FRB}

A white dwarf merger origin may also explain the repeating FRB in M81
\citep{Margalit2019, Kirsten2022,Kremer2021_FRB,Lu2022}. 
The detection of a single globular cluster FRB source out to the distance of M81 \citep[the enclosed volume of which contains roughly $1000$ total globular clusters;][]{Kremer2023_FRB} combined with the white dwarf merger rate of roughly $10^{-9}\,\rm{yr}^{-1}$ per cluster inferred from $N$-body simulations imply an active FRB lifetime of roughly $10^5-5\times10^6\,$yr (at $90\%$ confidence accounting for Poisson uncertainty associated with detection of a single event) is necessary for this source \citep{Lu2022,Kremer2023_FRB}.

The characteristic spin-down luminosity of a neutron star is
\begin{equation}
    \label{eq:lum_sd}
    \dot{E} \approx 10^{37}  \Bigg( \frac{\it{B}_{\rm{ns}}}{10^{11}\,\rm{G}}\Bigg)^{2} \Bigg( \frac{\it{P}_{\rm{ns}}}{10\,\rm{ms}}\Bigg)^{-4} \rm{erg\,s}^{-1}
\end{equation}
(see solid line contours in Figure~\ref{fig:PPdot}). The time-averaged isotropic equivalent luminosity inferred for the M81 FRB source is roughly $10^{29}f_{\rm{r}}^{-1}\,\rm{erg\,s}^{-1}$  \citep[e.g.,][]{Kremer2021_FRB} where $f_r$ is the uncertain efficiency factor for creating coherent radio emission. For $f_{\rm{r}}\gtrsim10^{-8}$ \citep[consistent with the Galactic FRB;][]{CHIME2020}, the spin-down luminosity is sufficient to power the M81 FRB source for the required $10^5-10^6\,$yr. 

Remarkably, this timescale and energy budget are consistent with the birth properties expected for neutron stars formed through white dwarf mergers (green points in Figure~\ref{fig:PPdot}). This hints that (i) white dwarf mergers may provide an avenue for formation of neutron stars capable of powering FRBs similar to the M81 source and (ii) given the discussion in Section~\ref{sec:NSs}, the four young pulsars in the Milky Way globular clusters may be descendants of such FRB sources.

If indeed such rotation-powered neutron stars emit FRBs, why have no such FRBs been detected from Galactic sources like the Crab pulsar which are formed at much higher rates (roughly $10^{-3}\,\rm{yr}^{-1}$ in the Milky Way)? Relatively short spin-down times for Crab-like objects combined with high beaming factors for FRB emission may provide a solution. Alternatively, the M81 FRB may instead be powered by magnetic activity \citep[e.g.,][]{BeloborodovLi2016} in a neutron star with $B \gtrsim 10^{14}\,$G \citep[e.g.,][]{Lu2022}. Such magnetar-level field strengths may be attained during collapse of white dwarf merger remnants if the field increases via dynamo processes \citep[e.g.,][]{GarciaBerro2012}, however if the magnetic fields are too high, the activity timescales become inconsistent with the ages required from white dwarf merger rates \citep{Kremer2021_FRB}. An alternative interpretation is that white dwarf mergers lead to multiple outcomes \citep[for example, depending on the mass ratio of the pair; e.g.,][]{Dan2014}: (i) magnetars capable of powering FRBs and (ii) neutron stars with $P\sim10$\,ms and $B\sim10^{12}\,$G that evolve into objects similar to the young pulsars observed in globular clusters. For a branching fraction of order unity, the relative rates of both the M81 FRB and the young pulsars could be accounted for. Regardless of a potential evolutionary connection between the M81 FRB and the observed young pulsars, the important role of white dwarf mergers in the formation of young neutron stars in globular clusters is clear.

Finally, the M81 FRB source is offset from the centre of its host cluster by roughly $2\,$pc \citep{Kirsten2022}, consistent with the offset expected for a neutron star receiving a natal kick of roughly $10-40\,$km\,${\rm s}^{-1}$ \citep{Kremer2023_FRB} as expected from ultra-stripped and/or electron-capture supernovae \citep[e.g.,][]{Tauris2015,Janka2017}. The large offset of the M81 FRB may appear to be in tension with the relatively small offsets of the four young globular cluster pulsars. However, unlike the FRB source which has an age $\lesssim10^6\,$yr, the four young pulsars are sufficiently old ($\gtrsim10^7\,$yr) to allow for two-body relaxation to return them to more central locations (Section~\ref{sec:offsets}).

\vspace{-0.7cm}
\section{Summary and Conclusions}
\label{sec:conclusion}

We have explored the formation of the four apparently young radio pulsars observed in Milky Way globular clusters through massive white dwarf mergers. We summarize our main findings below:

\begin{itemize}
    \item Neutron stars formed following collapse of white dwarf merger remnants are expected to have spin periods of roughly $10-100\,$ms and magnetic fields of roughly $10^{11}-10^{13}\,$G. These values are motivated by both observed properties of white dwarf merger remnants \citep[e.g.,][]{Ferrario2015,Caiazzo2021} and simulations of white dwarf mergers \citep[e.g.,][]{GarciaBerro2012,Schwab2021}. Such neutron stars will be observable as radio pulsars for roughly $10^7-10^8\,$yr before they spin down sufficiently to fall below the pulsar death line. The detection of four young pulsars in the Milky Way globular clusters with inferred ages of roughly $10^7-10^8\,$yr is consistent with expectations based on massive white dwarf merger rates from $N$-body cluster simulations \citet{Kremer2023_FRB}.
    \item Cluster simulations predict white dwarf mergers occur most frequently (by factors $\gtrsim 100$) in the dynamically oldest clusters that have undergone or are near core collapse. Notably, \textit{all four} of the young globular cluster pulsars are found in clusters of this type.
    \item Of the four, three of these young pulsars are single objects without a binary companion consistent with formation via merger of two white dwarfs. These three singles are all observed well within their host clusters' half-light radii, consistent with expectations from mass segregation. The exception is PSR J1718-19 which has both a binary companion and a relatively large cluster offset (roughly $6\,$pc). Both the binarity and offset may potentially be accounted for by a binary exchange encounter that led to large gravitational recoil.
    \item Finally, we argue that white dwarf mergers provide an avenue for formation of neutron stars capable of powering FRBs similar to the M81 source \citep{Kirsten2022}. This suggests a potential evolutionary connection between globular cluster FRB sources and the four young pulsars in the Milky Way.
\end{itemize}

An alternative interpretation is that these four pulsars are in fact not young, but instead were formed through disruption (via a binary dynamical encounter) during a low-mass X-ray binary phase. In this case the recycling process is halted, and these neutron stars appear as so-called partially-recycled objects \citep{VerbuntFreire2014}. X-ray binaries are well-known to be overabundant in the densest globular clusters \citep[e.g.,][]{Bahramian2013}, and thus may fit the trend shown in Figure~\ref{fig:rcrh}. Furthermore, such neutron star X-ray binaries may power some FRBs \citep[e.g.,][]{Sridhar2021}. Although the partial-recycling scenario cannot be ruled out to explain the apparently young pulsars, this scenario is unlikely to connect with the M81 FRB as periodicity \citep[a likely signature of binarity; e.g.,][]{Lyutikov2020} is not observed for this source \citep{Nimmo2023}.

In the coming years, radio instruments such as CHIME \citep{CHIME2018}, FAST \citep{Jiang2020}, the Square Kilometre Array \citep{SKA2009}, and DSA-2000 \citep{DSA2019} promise to uncover myriad new radio pulsars and FRBs. The M81 FRB specifically implies large numbers of analogous sources may be detectable in globular clusters of other nearby galaxies \citep{Kremer2023_FRB}. Additional detections of such FRB sources will yield further insight into the origin of these sources and their potential connection with Galactic radio pulsars.

\vspace{-0.6cm}

\section*{Acknowledgements}

We thank the anonymous referee, Ilaria Caiazzo, and Claire Ye for helpful discussions. Support for KK was provided by NASA through the NASA Hubble Fellowship grant HST-HF2-51510 awarded by the Space Telescope Science Institute, which is operated by the Association of Universities for Research in Astronomy, Inc., for NASA, under contract NAS5-26555. 

\vspace{-0.7cm}
\section*{Data Availability}

The data supporting this article are available on reasonable request to the corresponding author.


\vspace{-0.7cm}

\bibliographystyle{mnras}
\bibliography{mybib} 

\bsp	
\label{lastpage}
\end{document}